\documentclass[12pt]{iopart}

\usepackage{subfigure}

  \expandafter\let\csname equation*\endcsname\relax 
  \expandafter\let\csname endequation*\endcsname\relax 

\usepackage{amsmath}
\usepackage{dcolumn}
\usepackage{bm}
\usepackage{multirow}
\usepackage{floatflt} 
\usepackage{float} 

\usepackage{textcomp} 
\usepackage{iopams}
\usepackage{graphicx}
\usepackage{wasysym}
\usepackage{amssymb,amsfonts}
\usepackage{url}   

\newcommand{\Q}[1]{$Q^{'1}_{\text{#1}}$}
\newcommand{\QQ}[1]{$Q^{'2}_{\text{#1}}$}
\newcommand{\fbminus}[1]{\mbox{$f_{\beta^{e+p-n}_{\text{#1}}}$}}
\newcommand{\fbbarminus}[1]{\mbox{$f_{\beta^{\bar{e}+\bar{p}-\bar{n}}_{\text{#1}}}$}}
\newcommand{\fbplus}[1]{\mbox{$f_{\beta^{e+p+n}_{\text{#1}}}$}}
\newcommand{\fbbarplus}[1]{\mbox{$f_{\beta^{\bar{e}+\bar{p}+\bar{n}}_{\text{#1}}}$}}
\newcommand{\target}{$7 \cdot 10^{-13}$ }

\begin{document}

\title[Testing the UFF with Rb-Yb in VLBAI]{Testing the universality of free fall with rubidium and ytterbium in a very large baseline atom interferometer}

\author{J~Hartwig, S~Abend, C~Schubert, D~Schlippert, H~Ahlers, K~Posso-Trujillo, N~Gaaloul, W~Ertmer, and E~M~Rasel}

\address{Institut f\"ur Quantenoptik, Leibniz Universit\"at Hannover, Welfengarten 1, 30167 Hannover}
\ead{schubert@iqo.uni-hannover.de, gaaloul@iqo.uni-hannover.de}


\begin{abstract}
We propose a very long baseline atom interferometer test of Einstein's equivalence principle (EEP) with ytterbium and rubidium extending over 10\,m of free fall. In view of existing parametrizations of EEP violations, this choice of test masses significantly broadens the scope of atom interferometric EEP tests with respect to other performed or proposed tests by comparing two elements with high atomic numbers. In a first step, our experimental scheme will allow reaching an accuracy in the E\"otv\"os ratio of $7\cdot10^{-13}$. This achievement will constrain violation scenarios beyond our present knowledge and will represent an important milestone for exploring a variety of schemes for further improvements of the tests as outlined in the paper. We will discuss the technical realisation in the new infrastructure of the Hanover Institute of Technology (HITec) and give a short overview of the requirements to reach this accuracy. The experiment will demonstrate a variety of techniques which will be employed in future tests of EEP, high accuracy gravimetry and gravity-gradiometry. It includes operation of a force sensitive atom interferometer with an alkaline earth like element in free fall, beam splitting over macroscopic distances and novel source concepts.

\end{abstract}

\submitto{\NJP}


\maketitle
\section{Introduction}
\label{intro}

Einstein's equivalence principle (EEP) is  at the core of our understanding of gravitation and is  among the most important postulates of modern physics. It is under constant scrutiny since a violation of any of its pillars would lead to new physics beyond general relativity (GR) and would mark an important milestone in the search for a theory of everything (TOE). The EEP is comprised of three separate postulates: the Universality of Free Fall (UFF), Local Lorentz Invariance (LLI) and Local Position Invariance (LPI). Free fall experiments, as the one described in this letter, test the UFF by comparing the accelerations of two bodies of different internal structure and mass in a gravitational field. This inertial and gravitational mass equality is also known as the weak equivalence principle (WEP). To quantify a possible violation of the UFF it is common to normalise the acceleration difference between two test masses to the average local gravitational acceleration. This parametrization leads to the E\"otv\"os ratio defined by
$$\eta_{A,B}=2 \frac{g_{A}-g_{B}}{g_{A}+g_{B}},$$
with $g_{A,B}$ being the gravitational acceleration of test masses $A$ and $B$ respectively. The most straightforward way to do such a test is to directly measure the acceleration of two bodies in the same gravitational field. This class of tests is called Galilean and the most accurate to date was performed by comparing uranium and copper at a level of $10^{-10}$~\cite{Niebauer1987PRL}. The most accurate tests of the UFF were performed by the lunar laser ranging project (LLR), measuring the free fall of the moon and the earth in the gravitational field of the solar system. Since the UFF is a statement about the acting forces, not only Galilean type free fall experiments are performed to test it, but also force balance experiments with torsion balances. Torsion balances and LLR constrain possible violations of UFF to less than $10^{-13}$ in E\"otv\"os ratio~\cite{Williams2004PRL,Schlamminger2008PRL}. No violation was found so far. Future experiments with classical bodies are striving towards spaceborne platforms, to reduce the influence of external error source and allow measurements far beyond current state of the art~\cite{Nobili2012CQG,Touboul2012CQG}.\\
The use of atom interferometry broadens the field of test masses and allows an operation in the quantum regime. As such it is a complementary method to experiments with macroscopic bodies and will test aspects formerly inaccessible, such as violations linked to the coherence length of the test mass~\cite{Goklu2008CQG}, the possibility to employ cold atoms as accelerometers and clocks, and the possibility of spin-polarisation~\cite{Tarallo2014PRL}. A first measurement was performed by a device measuring gravity with a fountain of cold caesium atoms and comparing their fall rates to a commercial falling corner cube gravimeter at a level of $7\cdot 10^{-9}$~\cite{Peters1999Nat}. More recent experiments demonstrate tests of the UFF by using atom interferometry with two different quantum objects within the same device but do not yet reach the same precision. They are in part relying on two isotopes of the same species~\cite{Fray04PRL,Bonnin2013PRA,Tarallo2014PRL} but also on isotopes of two different elements~\cite{Schlippert2014PRL}. Especially tests with two isotopes want to take benefit from similarities for large noise suppression factors intrinsically arising
from the measurement’s arrangement. New experiments of both types are proposed to exceed the limits of current sensitivities, either on ground~\cite{Dickerson2013PRL,Dimopoulos2007PRL} or in micro-gravity environments~\cite{Rudolph2011MST,Geiger2011natcomm}, including the STE-QUEST space mission~\cite{Aguilera2014CQG}.\\
To employ this variety of test candidates in a precision experiment, a crucial point is the ability to trap both of the species not only simultaneously but rather in the same trap to have a well defined overlap of their initial positions and velocities. In this respect we propose quantum degenerate mixtures of rubidium and ytterbium for testing the UFF in a large scale device on ground.\\
In this paper we discuss the unique features of these mixtures that make them an ideal choice as test masses by calculating their violation parameters and comparing them to the ones used in other experiments and recent proposals. Focusing on the miscibility of different isotopes of these two elements, we will give a description on the source setup we aim for. Besides this description, we present possible scenarios for performing a UFF test with Bragg-type beam splitters. Along this we analyze noise contributions to the measured signal and estimate the performance of a test of the UFF to be \target in the E\"otv\"os ratio.
\section{Choice of test pairs}
\label{parameters}

As already mentioned the common way of quantifying an experiment testing the UFF is the E\"otv\"os ratio, which scales a measured differential acceleration to the strength of the local gravitational field comparing any abnormal composition based forces to the composition independent force. While this is a reasonable way to quantify the result of the performed measurement it does not take into account the specific kind of composition dependence in question. By just using the E\"otv\"os parameter as a tool for comparing two tests, an experiment with two spin polarized samples of the same isotope would not be treated different than a comparison between hydrogen and anti-hydrogen as proposed in~\cite{Hamilton2014PRL} while being fundamentally different. Taking the specific composition difference into account is part of the interpretation of the data and is strongly dependent on the model used to assess a possible violation theory. The use of extended wave functions for testing UFF opens the path to formerly unexplored theoretical models which are probing the quantum nature of matter and its interaction with space time~\cite{Goklu2008CQG}. While this is a vast field of study, we will focus on models which allow us a comparison to classical experiments. Specifically we asses the dilaton scenario~\cite{Damour2012CQG} and a scenario-independent scaling approach based on the standard model extension (SME)~\cite{Hohensee13PRL}. Atom interferometry can provide several new aspects different with respect to classical test masses as the test masses  are of high isotopic purity and the choices of test masses can be extended beyond non-magnetic, conducting solids which are typically used in torsion balances. \\
According to the dilation model~\cite{Damour2012CQG} a violation may be caused by forces acting differently on neutron and proton number. With the introduced effective charges \Q{A,B} and \QQ{A,B} calculated from the composition of a test particle a measurement of the E\"otv\"os ratio set bounds to the parameters $D_1$ and $D_2$ according to the formula 
\begin{equation}
\eta_{\text{A,B}}~\widetilde =~ D_1(\Delta Q^{'1}_{\text{A,B}})+D_2(\Delta Q^{'2}_{\text{A,B}})\text{.}
\end{equation}\\
A similar kind of parametrization can be given for the standard model extension~\cite{Hohensee13PRL}
\begin{equation}
\eta_{\text{A,B}}~\widetilde = ~\Delta f_{-n}+\Delta f_{+n}+\bar{\Delta f_{-n}}+\bar{\Delta f_{+n}}
\end{equation}
with the defined violation parameters for matter and anti-matter linked to neutron excess and total baryon number
\begin{equation}
\begin{aligned}
\Delta f_{-n} = f_{\beta^{e+p-n}_{\text{A}}}\beta^{e+p-n} - f_{\beta^{e+p-n}_{\text{B}}}\beta^{e+p-n}\\
\Delta f_{+n} = f_{\beta^{e+p+n}_{\text{A}}}\beta^{e+p+n} - f_{\beta^{e+p+n}_{\text{B}}}\beta^{e+p+n}\\
\bar{\Delta f_{-n}} = f_{\beta^{\bar{e}+\bar{p}-\bar{n}}_{\text{A}}}\beta^{\bar{e}+\bar{p}-\bar{n}} - f_{\beta^{\bar{e}+\bar{p}-\bar{n}}_{\text{B}}}\beta^{\bar{e}+\bar{p}-\bar{n}}\\
\bar{\Delta f_{+n}} = f_{\beta^{\bar{e}+\bar{p}+\bar{n}}_{\text{A}}}\beta^{\bar{e}+\bar{p}+\bar{n}} -f_{\beta^{\bar{e}+\bar{p}+\bar{n}}_{\text{B}}}\beta^{\bar{e}+\bar{p}+\bar{n}}\text{.}
\end{aligned}
\end{equation}
In both models larger absolute differences in the sensitivity factors of the employed test mass pair give rise to a larger signal in case of a violation of the UFF. Vice versa, an experimental determination of the E\"otv\"os ratio for such a test mass choice better constrains the existence of violations than tests performed with lower sensitivity factors for the same accuracy. Moreover, different test mass pairs probe different linear combinations of suspected violations linked to the neutron excess and the total baryon number of the test masses. In order to unambiguously determine the origin of a violation, a minimum of two test mass pairs needs to be employed. Interestingly, as shown in Ref.~\cite{Mueller2013proc}, even a test performed at a lower accuracy as compared to state of the art tests can further constrain possible violations, when the used test masses are significantly different to previously utilized ones. The sensitivity factors for different choices of test pairs are presented in table~\ref{violation}. For example, in comparison to Be-Ti the combination of ytterbium and rubidium isotopes is a factor of 2 more sensitive to baryon number related violations and even three orders of magnitude more sensitive in the parameter $\bar{\Delta f_{-n}}$.
 
\begin{table*}[h!]
\caption{Comparison of choices for test masses A and B employed in existing and planned tests of the UFF parametrized for violation scenarios with respect to their effective charges \Q{A,B}, \QQ{A,B}~and \fbplus{A,B}, \fbminus{A,B}, \fbbarminus{A,B}, \fbbarplus{A,B} calculated according to \cite{Damour2012CQG} and \cite{Hohensee13PRL}. Nuclide data is used from~\cite{Audi03} and for Ti a natural occurrence of isotopes is assumed~\cite{Laeter09}.}
\begin{tabular}{ c c c | c c  c c c c } \hline
  \multirow{2}{*}{A}& \multirow{2}{*}{B}&\multirow{2}{*}{Ref.} &\multicolumn{1}{c}{$\Delta$\Q{A,B}}&\multicolumn{1}{c}{$\Delta$\QQ{A,B}}& \multicolumn{1}{c}{$\Delta f_{-n}$} & \multicolumn{1}{c}{$\Delta f_{+n}$} & \multicolumn{1}{c}{$\bar{\Delta f_{-n}}$} & \multicolumn{1}{c}{$\bar{\Delta f_{+n}}$} \\
  &&&\multicolumn{1}{c}{$\cdot 10^4$}&\multicolumn{1}{c}{$\cdot 10^4$}&\multicolumn{1}{c}{$\cdot 10^2$}&\multicolumn{1}{c}{$\cdot 10^4$}&\multicolumn{1}{c}{$\cdot 10^5$}&\multicolumn{1}{c}{$\cdot 10^4$}\\
\hline
\textsuperscript{9}Be& Ti&\cite{Schlamminger2008PRL}  &-15.46 &-71.20& 1.48 &-4.16 & -0.24 &-16.24\\
Cu &\textsuperscript{238}U&\cite{Niebauer1987PRL} &-19.09 & -28.62 &-7.08& -8.31 &-89.89& -2.38\\
\textsuperscript{6}Li&\textsuperscript{7}Li &\cite{Hohensee2011JMO} &0.79& -10.07 &-7.26& 7.79 &-72.05& 5.82\\
\textsuperscript{85}Rb&\textsuperscript{87}Rb&\cite{Fray04PRL,Fray2009SSR,Bonnin13PRA} &0.84& -0.79 &-1.01& 1.81 &1.04& 1.67\\
\textsuperscript{87}Sr&\textsuperscript{88}Sr &\cite{Tarallo2014PRL} &0.42& -0.39 &-0.49& 2.04 &10.81& 1.85\\
\textsuperscript{39}K&\textsuperscript{87}Rb&\cite{Schlippert2014PRL}& -6.69& -23.69& -6.31& 1.90& -62.30& 0.64\\
\textsuperscript{87}Rb&\textsuperscript{170}Yb&[This work]& -12.87 &-13.92 &-1.36& -8.64 &86.00 &-5.46\\ \hline

\end{tabular}
\label{violation}
\end{table*}
\section{Atom interferometry in a 10~m atomic fountain}

\label{vlbai} 
The inertial sensitive interferometry with cold rubidium clouds is well covered by state-of-the-art experiments for measuring gravity~\cite{Hauth2013APB,Gillot2014Met}, gravity gradients \cite{Rosi2014Nat} and rotations~\cite{Tackmann2012NJP} as well as for measuring fundamental constants \cite{Bouchendira2011PRL}. Similarly laser-cooled ytterbium is by now very successfully utilized in optical clocks, especially optical lattice clocks~\cite{Hinkley2013Sci}. A key prerequisite to perform interferometry over long baselines is the preparation of a very narrow velocity distribution even beyond the ones of typical Bose-Einstein condensates which was already demonstrated for both species~\cite{Anderson1995Sci,Cornish2000PRL,Takasu2003PRL,Yamazaki2010PRL}. This can be reached by delta-kick cooling~\cite{Muntinga2013PRL,Kovachy2014Arx}. The facility we want to employ for a test of the UFF is the {\it VLBAI-Teststand} located at the new founded Hanover Institute for Technology (HITec)~\cite{HITEC_WEBPAGE}. This device will provide two experimental chambers for the preparation of atomic ensembles with two independent source chambers for a maximum flexibility in the choice of atomic species. A 10\,m ultra-high vacuum-tube with a magnetically shielded region of approximately 9\,m forms the baseline for an extended free fall. Since operation of the equivalence principle test only occurs in the magnetically shielded region we anticipate a free fall time of 1\,s and up to 2.6\,s if the atoms are launched. Assuming a measurement with $1\cdot 10^{5}$ ytterbium atoms and $2\cdot 10^{5}$ rubidium atoms produced in 10\,s, this leads to a shot noise limited performance of $1.6\cdot 10^{-10}\,\mathrm{Hz}^{-1/2}$ and $6.5\cdot 10^{-12}\,\mathrm{Hz}^{-1/2}$ in the E\"otv\"os ratio respectively. The second value relies on higher order beam splitters, as explained in chapter \ref{requirements_accuracy}.\\

\begin{figure}[t]
        \centering
        \begin{subfigure}[Mach Zehnder geometry]{
		\includegraphics[width=7cm]{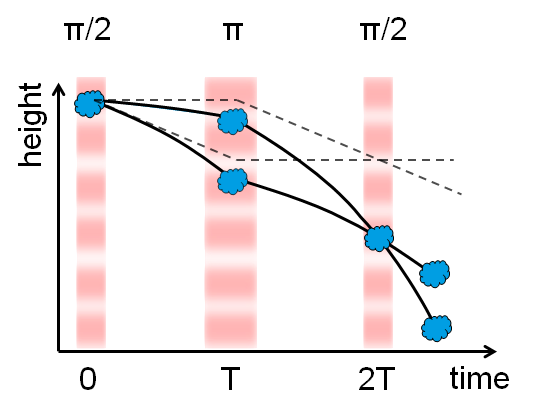}
		\label{machzehnderscheme}}
        \end{subfigure}
        \hspace{1 cm}
        \begin{subfigure}[Setup]{
        \includegraphics[width=2cm]{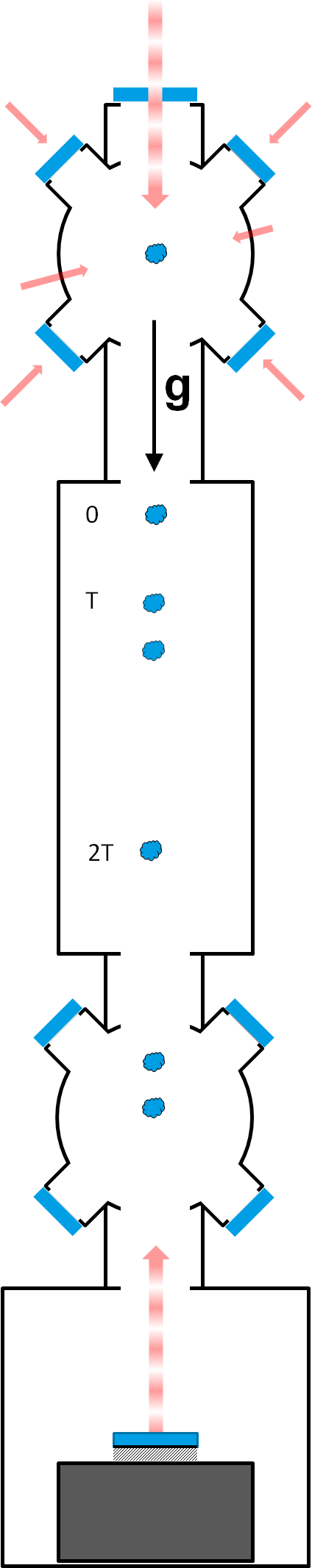}
        \label{experimentalscheme}}
        \end{subfigure}
        \caption{Mode of operation in Mach-Zehnder configuration and sketch of the experimental setup. Shown in \ref{experimentalscheme} is an operation in drop configuration.}
\end{figure}

\section{Concept for a dual species source of rubidium and ytterbium}
\label{source}
Mixtures of rubidium and ytterbium have been studied before in various experiments \cite{Munchow2011PCC,Baumer2011PRA} but were not yet used for precision interferometry. The construction of a dual species source capable of supporting an EEP test experiment faces a variety of challenges which are studied in the first phase of the experiment described in this work. A source has to fulfill the following characteristics:

\begin{itemize}
\item The clouds have to be able to be cooled down to quantum degeneracy  to fully exploit the long time of free fall achievable in the used infrastructure. Although this is relaxed by employing so called delta kick cooling, the efficiency of this process is strongly dependent on the initial temperature.
\item The initial collocation has to be very well known and controlled. To a certain degree this excludes isotope combination which are immiscible as discussed in chapter~\ref{mixtures}.
\item The initial velocity distribution of the two species has to be matched to a high degree to allow for differential suppression of systematic effects, like wave front curvature or residual rotations.
\item To achieve the target performance, $1\cdot 10^{5}$ ytterbium atoms and $2\cdot 10^{5}$ rubidium atoms have to be brought to degeneracy in less than 10\,s. If this performance is not reached, it will increase the time needed for integration, but is not prohibitive to the overall experiment.
\end{itemize}


\subsection{MOT Operation} Rubidium has two stable isotopes with mass numbers 87 and 85, both are bosonic and can be brought to degeneracy with common methods \cite{Anderson1995Sci,Cornish2000PRL}. Since both are also natural abundant and can be cooled similar well by standard laser cooling techniques, the specific decision for a rubidium species will be taken based on the miscibility with the ytterbium isotopes. The widely spread method for the preparation of rubidium ensembles is laser-cooling on the $5^2S_{1/2}$-$5^2P_{3/2}$ transition with a subsequent optical molasses step for achieving sub-Doppler temperatures down to approximately $2\,\mu$K. With a combination of a multi-layer atom chip allowing for an efficient transfer of laser cooled atoms to a magnetic trap and a 2D$^+$-MOT, quantum degenerated ensembles with $4\cdot10^{5}$ rubidium atoms were produced in 1.6\,s~\cite{Rudolph2015arXiv}.\\ 
With in total five bosonic and two fermionic stable isotopes that have all been brought to quantum degeneracy before \cite{Takasu2003PRL,Yamazaki2010PRL}, ytterbium offers a variety of choices for test masses as seen in table~\ref{yb_isotopes}. The bosonic isotopes have no hyperfine splitting and therefore a very low magnetic sensitivity compared to rubidium for example \cite{Taichenachev2006PRL}. While this is  beneficial to counteract systematic effects, the missing possibility to drive Raman-transitions between the hyperfine states is limiting the implementation scenarios. Ytterbium, an alkaline earth like element, offers the possibility to perform narrow-line cooling on the inter-combination transition $^1S_0$-$^3P_1$ with a Doppler-temperature of $T_D=4.4\,\mu$K. Due to a low vapor pressure one has to face the challenge to pre-cool the hot source for efficient MOT operation. The common method is the use of a Zeeman-slower with a transversal cooling stage at the singlet transition $^1S_0$-$^1P_1$ \cite{Miranda2012PRA}. Another comparably new option is the use of 2D-MOT at same transition \cite{Dorscher2013RSI}. Experimentally loading rates of $6\cdot 10^7$ $^{174}$Yb atoms per second have been achieved by both methods. The 2D-MOT seems preferable over the Zeeman-slower setup in terms of vacuum quality in the main chamber due to the use of differential pumping stages and offers higher scalability with available laser power at 398.9\,nm.
 
\begin{table*}[h!]
\caption{Stable isotopes of ytterbium and their relative natural abundance~\cite{Lide2008CRC} in $\%$, character of spin-statistic, intra-species scattering length~\cite{Kitagawa2008}, inter-species scattering length with $^{87}$Rb in $a_0$~\cite{Borkowski2013arXiv}, as well as isotope-shift relative to $^{174}$Yb of the relevant cooling transitions in MHz.}
\begin{tabular}{ c | c c c c c c c } \hline
Isotope & Abund. & Spin st. &$a_{Yb/Yb}$ & $a_{Yb/Rb}$ & $J$ & $^1S_0$-$^3P_1$ & $^1S_0$-$^1P_1$\\
\hline
$^{168}$Yb & 0.13 & boson & $252 \pm 3$ & $39.2 \pm 1.6$ & & 3655 & 1887.4\\ 
$^{170}$Yb & 3.05 & boson & $64 \pm 2$ & $-11.5 \pm 2.5$ & & 2287 & 1192.4\\ 
$^{171}$Yb & 14.3 & fermion & $-2.8 \pm 3.6$& $58.9 \pm 4.4$ &(1/2-1/2)& -2132& 1153.7\\ 
& & & & &(1/2-3/2)& 3805 & 832.4\\ 
$^{172}$Yb & 21.9 & boson & $-599 \pm 64$ & $-161 \pm 11$ && 1000 & 1887.4\\ 
$^{173}$Yb & 16.1 & fermion & $199 \pm 2$ & $626 \pm 88$ &(5/2-5/2)& 2312 & -253.4 \\ 
 &  &  &  &  &(5/2-7/2)& -2386 & 588\\ 
 &  &  &  &  &(5/2-3/2)& 3806 & 516\\ 
$^{174}$Yb & 31.8 & boson & $105 \pm 2$ & $880 \pm 120$ & & 0 & 0\\ 
$^{176}$Yb & 12.7 & boson & $-24 \pm 4$ & $216.8 \pm 4.7$ & & -955 & -509.3\\  \hline
\end{tabular}
\label{yb_isotopes}
\end{table*}

\subsection{Trapping and evaporation} 
\begin{figure}
\centering
\begin{subfigure}[Scalar polarizability $^1S_0$]{
  \includegraphics[width=0.45\textwidth]{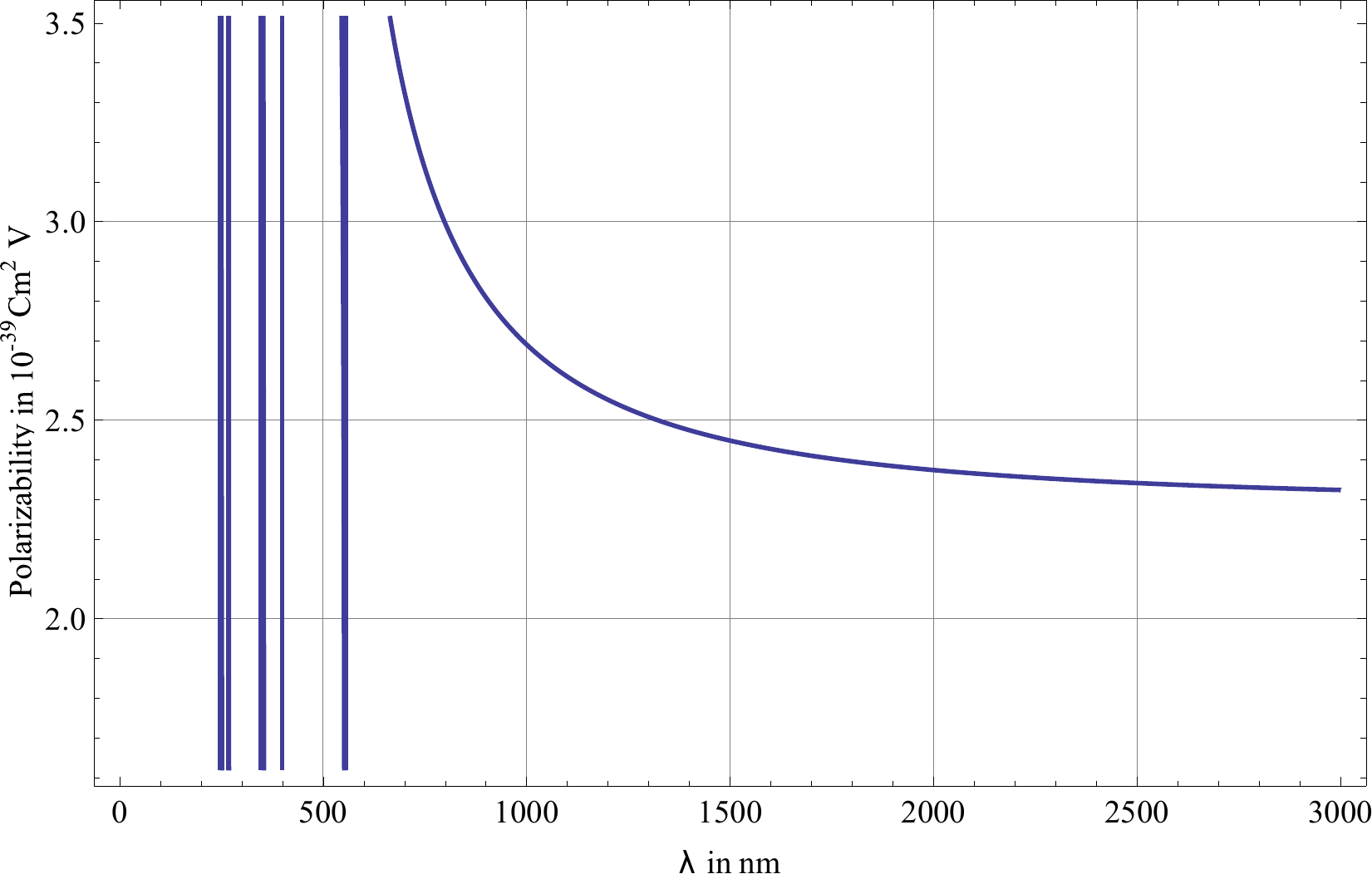}
  \label{scPol-S}}
\end{subfigure}
\begin{subfigure}[Scalar polarizability $^3P_1$]{
  \includegraphics[width=0.45\textwidth]{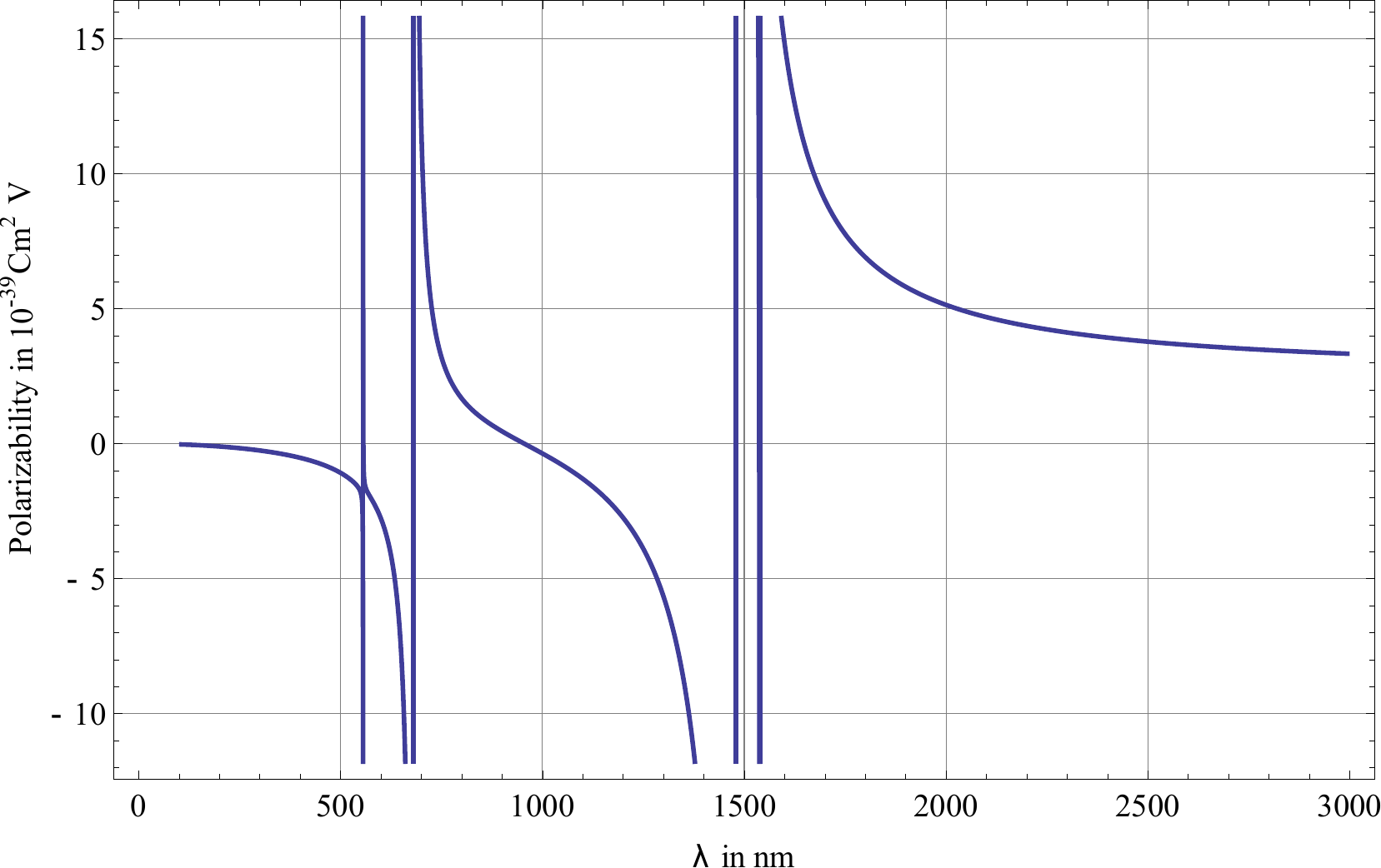}
  \label{scPol-P}}
\end{subfigure}
\begin{subfigure}[Differential scalar polarizability $^1S_0$-$^3P_1$]{
  \includegraphics[width=0.45\textwidth]{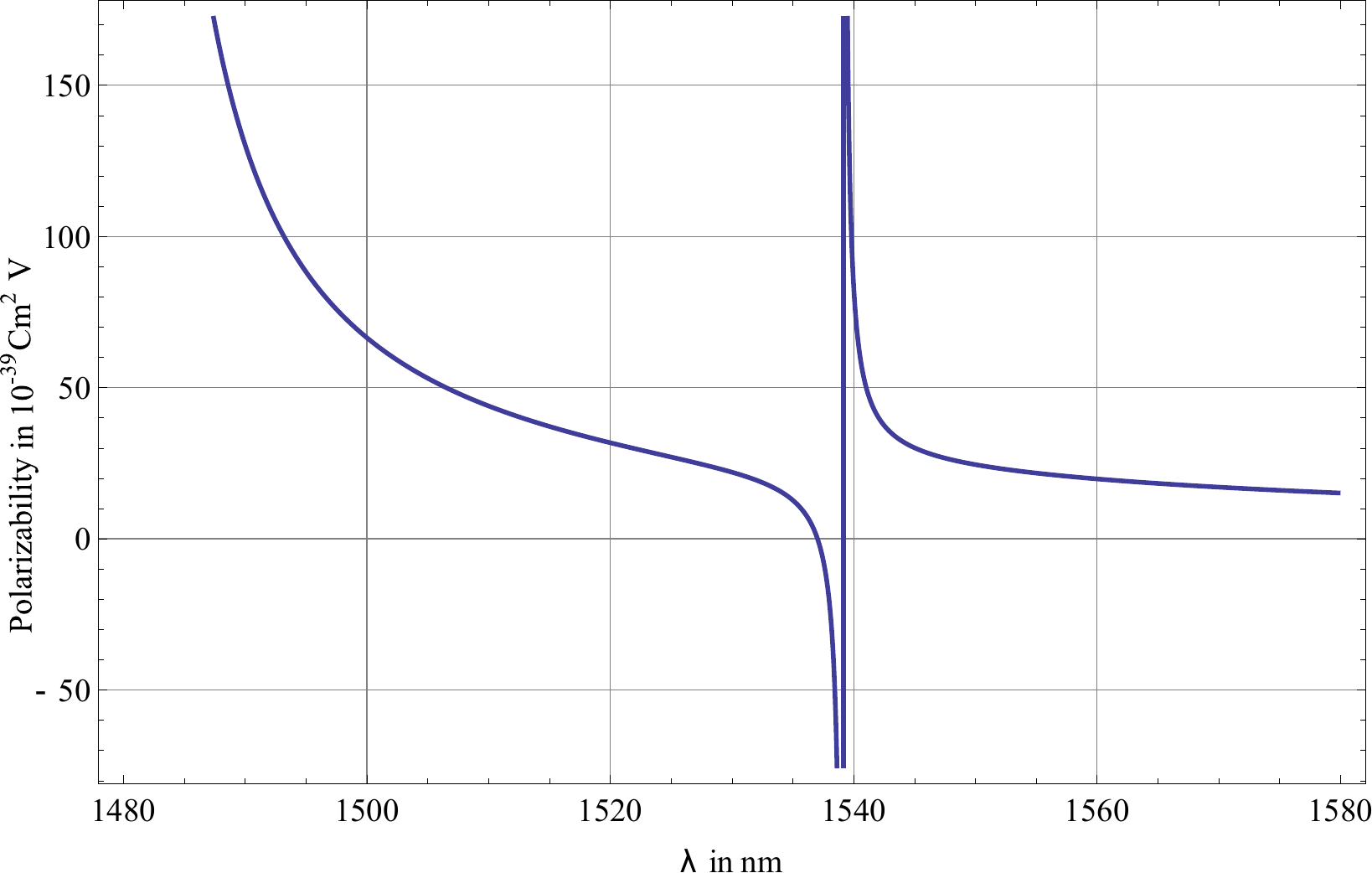}
  \label{diffPo-SP}}
\end{subfigure}
\begin{subfigure}[Differential AC-Stark shift]{
  \includegraphics[width=0.45\textwidth]{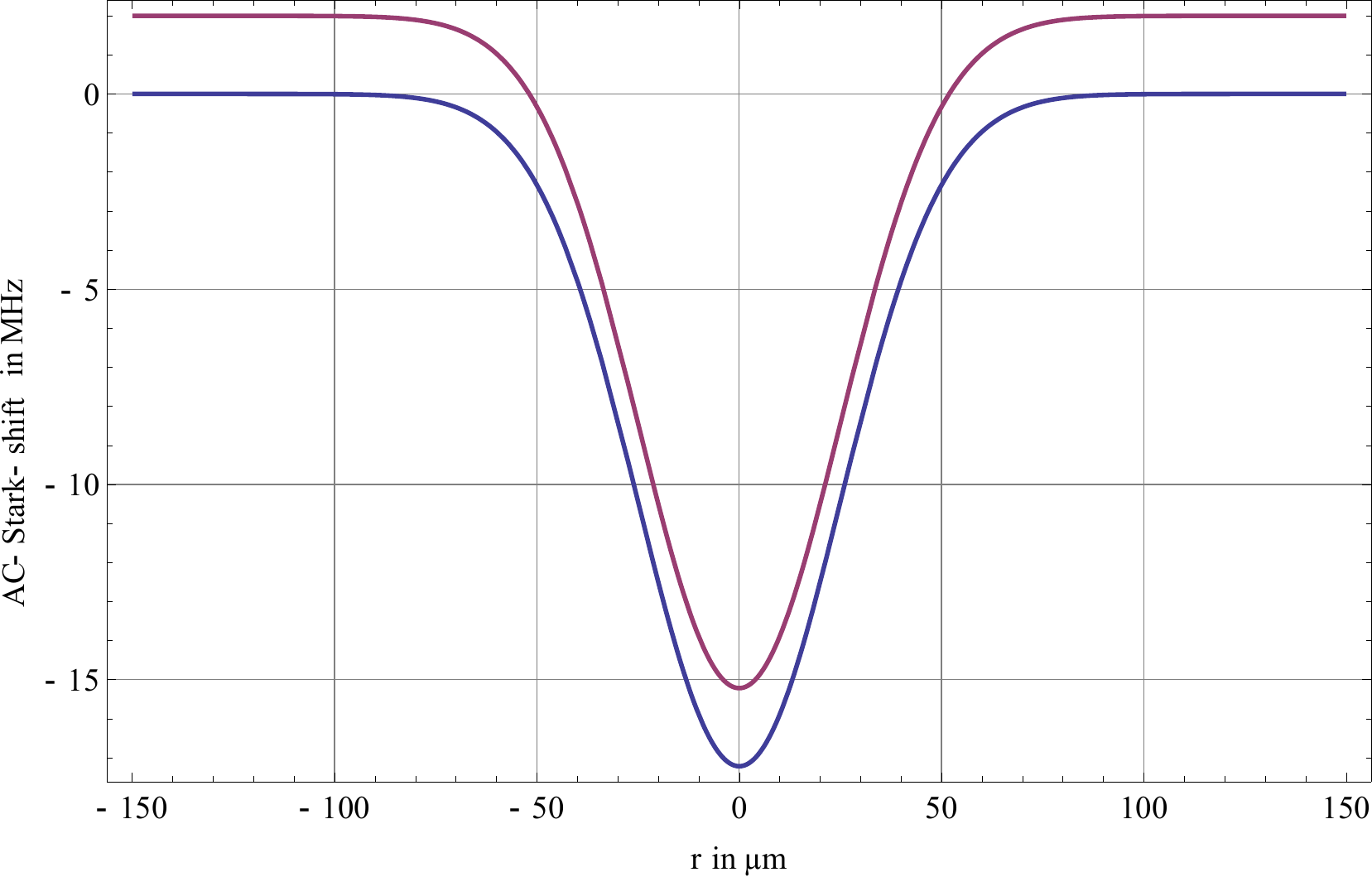}
  \label{AC-Stark}}
\end{subfigure}
\caption{Scalar polarisability and effective AC-Stark shift. The upper curves \ref{scPol-S} and \ref{scPol-P} show the laser wavelength dependent scalar polarisability of the states in the transition used for the intercombination line cooling. The lower curves show in \ref{diffPo-SP} the differential polarisability and in \ref{AC-Stark} the resulting differential AC Stark shift imposed on the intercombination line by a 1960\,nm ODT with 100\,W, a 50\,$\mu$m waist and using an additional 8.84\,mW dressing beam with 1\,GHz blue detuned to the transition.}
\label{polarsim} 
\end{figure}

Since we aim for a combined trap of both species, magnetic traps are not an option for the magnetically not trappable ytterbium. As a result a far detuned optical dipole-trap in the mid-infrared will be used as a common trap. Figure~\ref{polarsim} shows the scalar polarisability at a certain wavelength with respect to the inter-combination MOT for ytterbium. The differential polarisability shows mainly two remarkable results: Ytterbium is not trapped at $1\,\mu$m and there is a zero-crossing close to $1.5\,\mu$m, that would potentially allow for AC-Stark shift compensated dipole-trap. A more conservative and less demanding solution would be the use of a dipole-trap beyond the zero-crossing for example at 1960\,nm. To compensate AC-Stark shift dispersion over the cloud, which would be large due to the narrow linewidth of the transition  a low-intensity blue detuned compensation beam can be used \cite{Kaplan2002PRA} with a detuning of $\Delta_{\text{comp.}} = 2\pi\cdot1$\,GHz and a power of $I_{\text{comp.}} = 8.84$\,mW. The Bose-Einstein condensation in a single beam dipole-trap at this wavelength for $^{87}$Rb was already shown in a weak hybrid trap configuration in~\cite{Zaiser11PRA}. Therefore, a 1960\,nm trap appears to be an ideal solution and lasers with output powers up to 100\,W are available.

\subsection{Dual species loading sequence} The cycle time of the experiment will be limited by smaller loading rates of the ytterbium, even with the use of a 2D$^{+}$-MOT and the expected increase in flux, due to the use of higher laser power. In addition the $^1S_0$-$^1P_1$ transition cannot be driven together with the rubidium cooling transition $5^2S_{1/2}$-$5^2P_{3/2}$, since the ionization energy of the upper state of rubidium is 2.59\,eV that corresponds to 478.7\,nm. Therefore the dual species sequence will first completely undergo the loading steps for cooling and trapping ytterbium into the dipole trap before we start the fast loading of the rubidium MOT. To avoid losses due to collisions at this stage of the experiment it is possible to shift the center of the rubidium MOT against the dipole trap via adjusting the magnetic field gradient before both isotopes are co-located inside the dipole trap.

\subsection{Species miscibility and dynamical evolution}
\label{mixtures}

\begin{figure}[t]
        \centering
        \begin{subfigure}[$^{168}$Yb-$^{87}$Rb]{
		\includegraphics[width=0.4\textwidth]{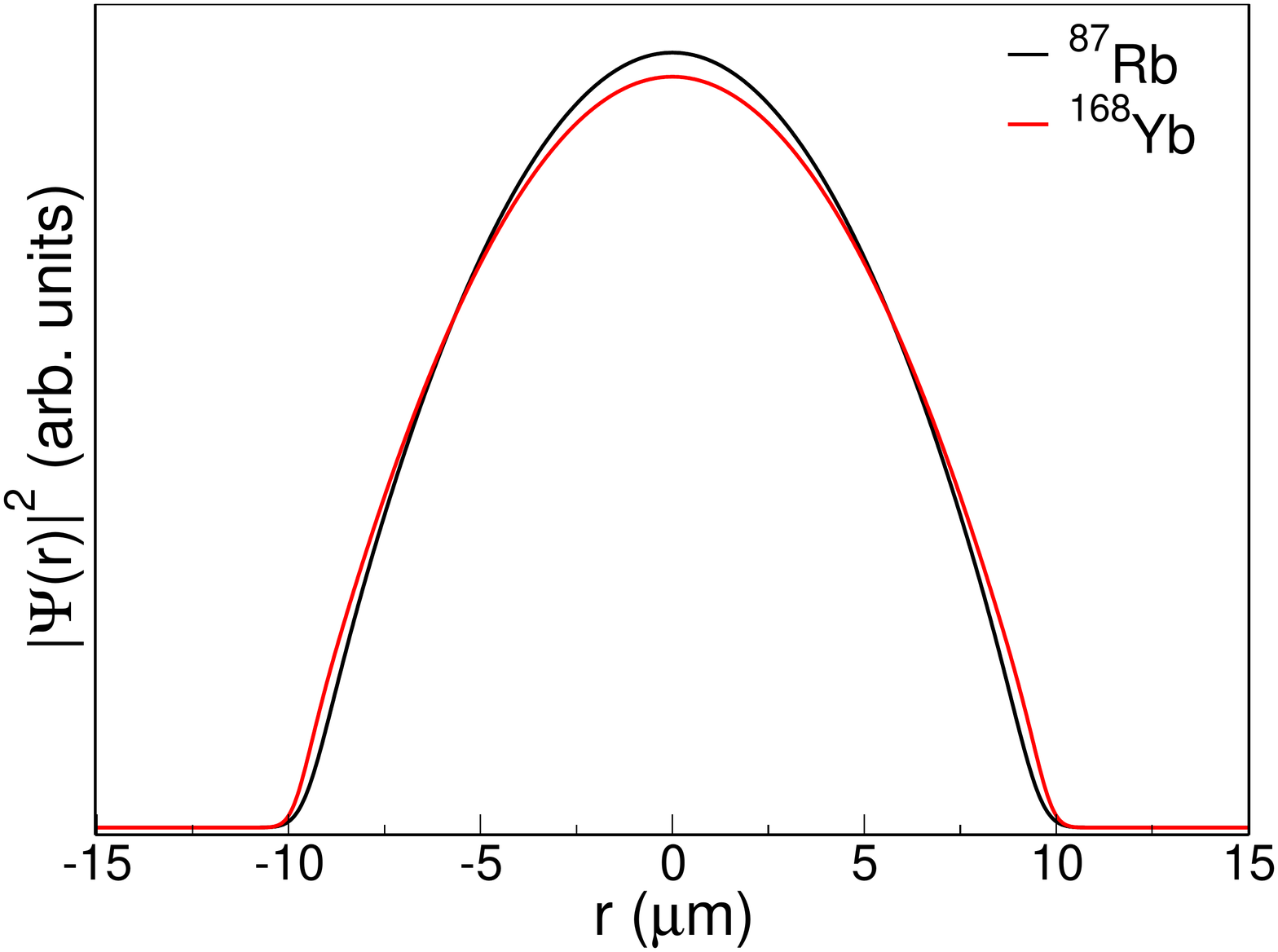}
		\label{miscibilitya}}
        \end{subfigure}
        \hspace{1cm}
        \begin{subfigure}[$^{170}$Yb-$^{87}$Rb]{
        \includegraphics[width=0.4\textwidth]{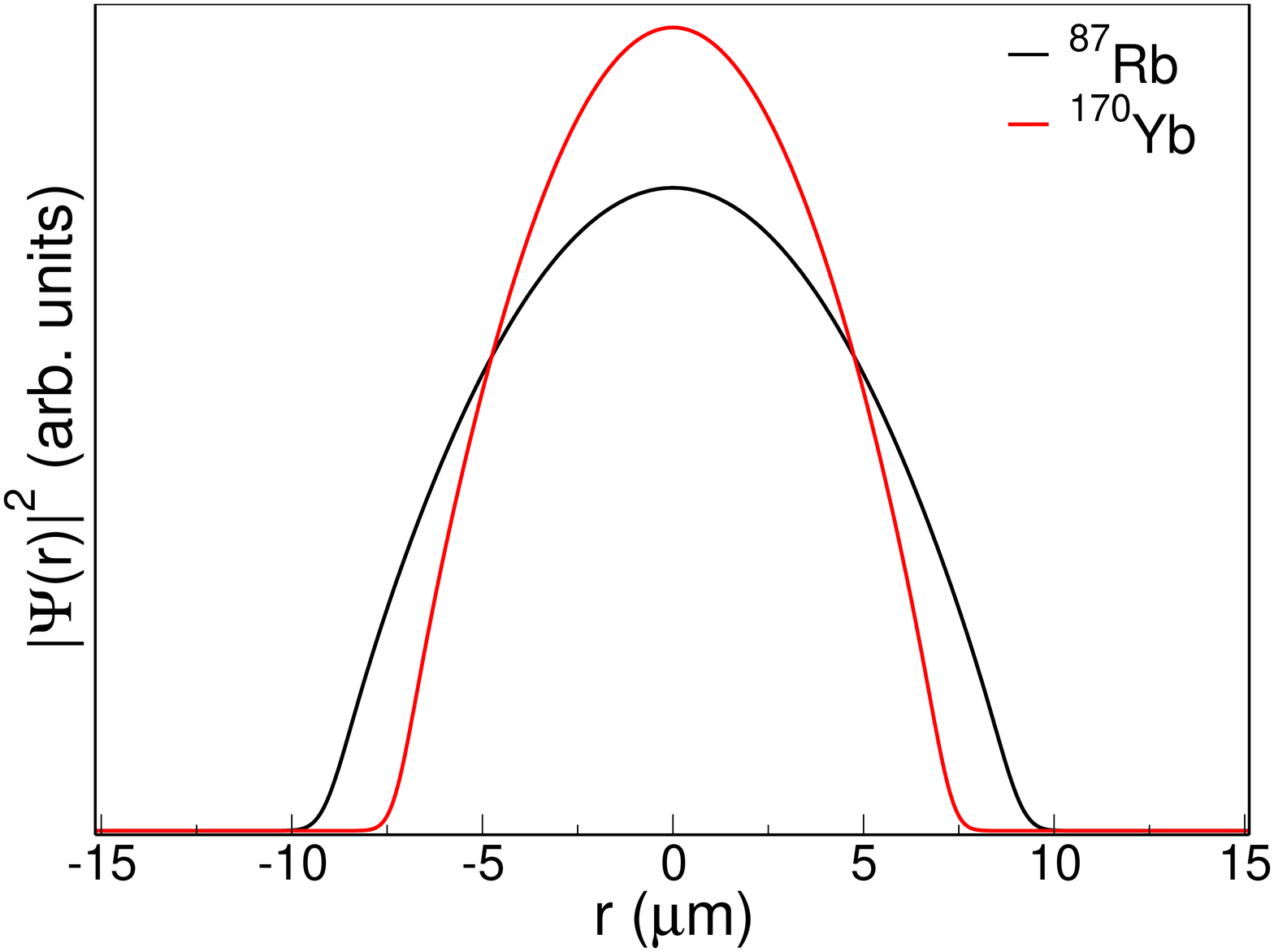}
        \label{miscibilityb}}
        \end{subfigure}
\caption{Density plots of the ground states of the $^{170}$Yb/$^{168}$Yb and $^{87}$Rb mixtures. For each pair mixture, the wave functions are computed solving the Gross-Pitaevkii equation in 3D including the intra-species interactions of the two isotopes and the inter-species one with $^{87}$Rb.
The magnitudes of these interactions are the same shown in table~\ref{yb_isotopes}. We assume that each mixture is confined by the same external trap with frequencies solely differing due to the mass difference. The trapping frequencies are $2\pi\cdot 88$\,Hz for Rb and $2\pi\cdot 67$\,Hz for Yb. In both cases, a symmetric mixture ground state is found illustrating the miscibility of the two pairs without further tuning of external optical or magnetic parameters (Feshbach for example).}
\label{miscibilityplot} 
\end{figure}

This ability to cool non-magnetic ytterbium isotopes to quantum degeneracy inside the 2\,$\mu$m dipole trap via evaporation without additional effort is a key motivation for our choice. Fermionic isotopes are not considered in this study since degenerate Fermi gases are large and expand with higher rates than BECs, which is an important parameter for long baseline interferometry. They might nevertheless be interesting for future tests and the device is designed to keep this option open. As table 2 shows, we are left with five bosonic isotopes where two of them, $^{172}$Yb and $^{176}$Yb, have negative intra-species scattering length. They would require a more complex experimental design including the manipulation of an optical Feshbach resonance to reach degeneracy. $^{174}$Yb is the most abundant isotope which was already condensed~\cite{Yamazaki2010PRL}. Nevertheless, due to the repulsive collisions to $^{87}$Rb (inter-species scattering lengths of $(880\pm 120)\,\textit{a}_0$), a binary mixture will not be stable due to three-body losses.
For all the reasons stated above, we focus our investigations on $^{168}$Yb, $^{170}$Yb and possible mixtures with $^{87}$Rb. Unfortunately, $^{168}$Yb and  $^{170}$Yb are the least abundant isotopes making loading rates significantly low which constrains the cycling rate in the order of tens of seconds unless they are enriched. The $^{168}$Yb -$^{87}$Rb  mixture features an inter-species positive scattering length of $39.2\pm 1.6\,\textit{a}_0$ meaning that this Yb isotope can be sympathetically cooled by $^{87}$Rb atoms. 
As shown in our systematics study in section \ref{requirements_accuracy}, the separation between the two components of a  binary mixture has a dramatic effect on the performance of the UFF test. Therefore, quantum miscibility cannot be neglected in this density regime. Indeed, if the interspecies repulsion exceeds the miscibility threshold~\cite{Papp2008PRL}, the two atomic clouds spatially separate to minimize the interaction energy. This immiscible state is a hindrance for optimising the overlap of the centre of mass of the two wave packets fed into the interferometer for  comparison. This makes it necessary to carefully check for the proposed isotopes if they can be prepared in overlapping pairs of spherical symmetry. We therefore solve a system of 3D-coupled Gross-Pitaevskii equations describing the ground state of the mixture \cite{Ho1996PRL}. The results of these simulations are shown in figure \ref{miscibilityplot}.

The calculations confirm the miscibility of $^{87}$Rb with the two Yb isotopes considered making it a suitable candidate for an UFF test. In contrast, the combination of $^{168}$Yb with $^{170}$Yb builds up a symmetric shell structure. These binary states numerically found are susceptible to and deformable by external fields (magnetic forces, gravitational sag, etc.) present in the science chamber. Therefore, this mixture is not considered for dynamics and systematics. \\
In order to reduce systematic errors of the atom interferometric comparison and allow for an extended interrogation time, it is crucial to reduce the size of the atomic samples. In the proposed facility, few seconds of free fall or launch time are used to reach the target accuracy of the UFF test. It is clear that thermal ensembles would reach very large sizes at these time scales. This motivates the use of degenerate matter waves characterized by a slow expansion. The state-of-the art in slowing down the expansion of BECs improved dramatically with the use of delta-kick cooling (DKC) techniques \cite{Muntinga2013PRL,Dickerson2013PRL}. In recent experiments with a comparable baseline \cite{Kovachy2014Arx}, it was experimentally demonstrated that the expansion energy of a degenerate  $^{87}$Rb ensemble could be restricted to only few tens of pK in 2D. We anticipated such records when proposing space missions with more than 10\,s of free evolution time~\cite{Aguilera2014CQG} of a mixture of a $^{87}$Rb / $^{85}$Rb condensates. \\
The DKC manipulation~\cite{Chu1986pro} consists in collimating matter waves by suddenly reducing the frequency of the initial trap holding the atoms and cutting it when all atoms reach the turning points of the trap walls (at t$_p$/4, where t$_p$ is the trap period). The same result is expected by re-pulsing the initial trap after switching it off for some free expansion time. A substantial part of the atoms kinetic energy is absorbed by this process leading to a slowed expansion. The analogy with light beams collimation often led to label this manipulation as an atomic lens. We anticipate the use of a double lens to match the expansion rates of $^{87}$Rb an $^{170}$Yb. This match is mandatory to mitigate errors related to residual wave front curvatures and relaxes the requirements on the initial collimation and retro reflection mirror planarity.
\subsection{Interferometer sequence}
\label{sequence}
As described earlier, performing an UFF-test is equivalent to a simultaneous measurement of the gravitational acceleration $g_{A,B}$ acting on the two test masses. To perform this measurement with atoms a sequence of light pulses has to be applied to interrogate them with respect to a common reference mirror which acts as a phase front reference. The most prominent configuration for inertial sensitive atom interferometry is the Mach-Zehnder-type $\pi/2-\pi-\pi/2$ sequence with a time $T$ of free evolution in between each of the pulses. Two different modes of operation can be distinguished: (i) dropping atoms from a source on the top of the device and (ii) launching atoms onto a parabolic trajectory from a source at the bottom of the device. While the first mode is characterized by a good control over the initial conditions at free evolution times of $2T=1-1.3$\,s at a baseline of roughly 9\,m, the second one offers the perspective to increase the overall length of the interferometer up to $2T=2.6$\,s. Launching over approximately 10~m was already demonstrated for rubidium in an accelerated optical lattice by coherently transferring a large number of photons at a decent efficiency \cite{Dickerson2013PRL} and appears also realizable for ytterbium with similar parameters. Nevertheless, this fountain mode requires a well controlled launching velocity of both test masses.

\subsection{Beam splitting and match of scaling factor.}
A major limitation for inertial measurements with atom interferometers is seismic noise which scales similar to the acceleration signal with $T^2$ and thus limits the maximum time of interferometry where the signal to noise ratio is still improving. When using a common mirror for a differential measurement, as planned for this experiment, the seismic noise for both interferometers is common and thus suppressed in the difference signal~\cite{Varoquaux2009NJP,Chen2014PRA}.
To fully benefit from the non magnetic properties of the ytterbium $^1S_0$ state and allow for higher order beam splitting we plan to use Bragg type beam splitters, coupling momentum states of the respective ground states. The used off resonant transitions are the $^1S_0$-$^1P_1$ transition for ytterbium at 399\,nm and the $5^2S_{1/2}$-$5^2P_{3/2}$ transition for rubidium at 780\,nm. 
The suppression factor depends on the match of the scaling factor $kT^2$, with the effective wave vectors $k$,  and of the sensitivity function  which is itself dependent on the timing of the interferometer pulse sequence. The basic approach is to match the scaling factors by tuning the interferometry time $T$ for each species individually~\cite{Varoquaux2009NJP}. This will lead to a small difference in the frequency response of the two interferometers and will not properly suppress contributions scaling differently with $T$ but allows for a simple data analysis scheme.\\
In the case of mismatched effective wave vectors and same pulse timing, the phase frequency response is similar between the two species but rescaled according to the appropriate wave vector. As long as the resulting phase noise is smaller than 1\,rad the phase information can still be fully recovered by weighting the results with the wave vector ratio. An analysis of this case can be found in~\cite{Chen2014PRA}. Even in the case of noise above $\pi$ most of the information can be recovered at the cost of signal to noise ratio. In the case of higher common noise contributions the resulting 2$\pi$ ambiguity can be fully resolved by operating an additional classical sensor \cite{Geiger2011natcomm}. Another option is to adapt the model used for data interpretation and recover at least some level of suppression by fitting an appropriate probability distribution.

\begin{table}[hp]
 \caption{Requirements to reach the stated uncertainty in $\eta$ in different configurations. 1) Assuming $T_{zz}=-2.91\cdot 10^{-6}~s^{-2}$, $|T_{xx}|=|T_{yy}|=|T_{zz}|/2$, and $T_{zzz}=\delta_{z}T_{zz}$. 2) Assuming a counter rotation $\Omega_{c}$ with quality $(\Omega_{c}-\Omega_{y})/\Omega_{y}$, $\Omega_{y}=57.5\,\mu$rad/s. 3) Waist after collimation lens w$_0$, ideal distance lens - fiber 40\,cm, assumed distance 40\,cm-df.}
 \begin{tabular}{p{0.3\textwidth}p{0.18\textwidth}p{0.18\textwidth}p{0.18\textwidth}}\hline
   Error source & Initial & Intermediate & Advanced \\ \hline
  Free evolution time & 500\,ms, 505.7\,ms & 500\,ms, 500\,ms  & 1300\,ms, 1300\,ms \\
  Effective wave vectors k$_{Rb}$, k$_{Yb}$ & \multicolumn{2}{c}{2$\cdot4\pi$/(780\,nm), 4$\pi$/(399\,nm)} & 8$\cdot4\pi$/(780\,nm), 4$\cdot4\pi$/(399\,nm) \\
  - Relative uncertainties & \multicolumn{3}{c}{$\pm10^{-8}$ in individual k,} \\
   & \multicolumn{3}{c}{$\pm10^{-14}$ in the ratio $(k_{Yb}-k_{Rb})/(k_{Yb}+k_{Rb})$} \\
  Rabi frequency match & \multicolumn{3}{c}{1\,\%} \\ \hline
  Common velocity v$_{0}$ & \multicolumn{2}{c}{-3\,m/s $(\pm 30\,\mu$m/s)}  & 12.75\,m/s $(\pm 0.1275\,\mu$m/s) \\
  Differential position & $1\,\mu$m & $1\,\mu$m$\pm0.1\,\mu$m & $1\,\mu$m$\pm10$\,nm \\
  Differential velocity & $1\,\mu$m/s & $0.1\,\mu$m/s & $31\,\mu$m$\pm10$\,nm/s \\ \hline
  Grav. gradients - $T_{zz}$ $^{1)}$ & $3\cdot10^{-6}$ $(\pm1.5\cdot10^{-10})$\,s$^{-2}$ & $3\cdot10^{-6}$ $(\pm3\cdot10^{-7})$\,s$^{-2}$ & $3\cdot10^{-6}$ $(\pm1\cdot10^{-9})$\,s$^{-2}$ \\
  Grav. grad. - $T_{zzz}$ & $<5\cdot10^{-9}$ $(\pm3.3\cdot10^{-11})$\,m$^{-1}$\,$s^{-2}$ & $<5\cdot10^{-9}$ $(\pm3.3\cdot10^{-11})$\,m$^{-1}$\,$s^{-2}$ & $<5\cdot10^{-9}$ $(\pm10^{-11})$\,m$^{-1}$\,$s^{-2}$\\ 
  Acceleration compensation via frequency scan to & $10^{-7}$\,m\,s$^{-2}$ & $5\cdot10^{-8}$\,m\,s$^{-2}$ & $10^{-8}$\,m\,s$^{-2}$ \\ \hline
  Counter rotation quality $^{2)}$ & \multicolumn{3}{c}{2\,\%}  \\
  Rotation $\Omega_x$ & \multicolumn{3}{c}{1\,$\mu$rad/s}  \\
  Rotation $\Omega_z$ & \multicolumn{3}{c}{44.4\,$\mu$rad/s}  \\ \hline
  Inside shield &  &  &  \\
  - Magnetic offset field B$_0$ & 25\,mG & 25\,mG & 25\,mG \\
  - Magnetic field gradient $\delta$B & 50\,$\mu$G/m  & 15\,$\mu$G/m & 1.5\,$\mu$G/m \\ 
  Inside source chamber &  &  &  \\
  after release B$_0$, $\delta$B & \multicolumn{3}{c}{$<1$\,G, $<0.1$\,G/m}\\ \hline
  Collimation df, w$_0$ $^{3)}$ & 100\,$\mu$m, 2.3\,cm & \multicolumn{2}{c}{100\,$\mu$m, 3\,cm}  \\
  Mirror quality & \multicolumn{3}{c}{$\lambda/20$} \\
  Initial sample radius & \multicolumn{3}{c}{300\,$\mu$m}  \\
  Effective temperature & 25.6\,nK($\pm1\,\%$), 50\,nK($\pm1\,\%$) & 2.5\,nK($\pm1\,\%$), 5\,nK($\pm1\,\%$) & 256\,pK($\pm1\,\%$), 500\,pK($\pm1\,\%$) \\ \hline
  Atom numbers & \multicolumn{3}{c}{2$\cdot$10$^{5}$ $(\pm 1\,\%)$, 10$^{5}$ $(\pm 1\,\%)$}  \\
  Scattering lengths & \multicolumn{3}{c}{(100.4$\pm0.1$)\,a$_0$\cite{van2002PRL}, (64$\pm2$)\,a$_0$, (11.5$\pm2.5$)\,a$_0$} \\
  Beam splitting accuracy & 0.01 & 0.001 & 0.001 \\ \hline
 \end{tabular}
 \label{tab:error_budget_reqs}
 \end{table}

\section{Requirements and error budget}
\label{requirements_accuracy}
\begin{table}[tb]
 \caption{Contributions of the different error sources to the uncertainty in $\eta$ in different configurations. 1) Requires back correction via knowledge of g, $T_{zz}$, $T_{zzz}$, and $\Omega_{y}$}
 \begin{tabular}{p{0.3\textwidth}p{0.18\textwidth}p{0.18\textwidth}p{0.18\textwidth}}\hline
   Error source & Initial & Intermediate & Advanced \\
  u$_{\eta}$ & in $10^{-12}$ & $10^{-13}$ & $10^{-14}$ \\ \hline
  Gravity gradient + position overlap & 0.3 & 0.3 & 0.3 \\
  Gravity gradient + velocity overlap & 0.15 & 0.15 & 0.4 \\
  Gravity gradient + g, v$_{0}$ & 0.15 & 0.15 & 0.15 \\
  Coriolis x & 0.23 & 0.23 & 0.23 \\
  Coriolis y & 0.2 & 0.2 & 0.2 \\
  Other terms $^{1)}$ & 1 & 1 & 1 \\\hline
  Magnetic fields & 0.3 & 1 & 1 \\ \hline
  Wave fronts & 5.1 & 5.2 & 5.7 \\ \hline
  Mean field & 1.3 & 3.6 & 3.9 \\ \hline
  Sum & 5.7 & 6.7 & 7.4 \\ \hline
 \end{tabular}
 \label{tab:error_budget_table}
 \end{table} This chapter summarizes the requirements on experimental and environmental parameters to restrict statistical and systematic errors. These requirements are partly relaxed compared to single species gravimetry measurements~\cite{Louchet2011NJP,LeGouet2008APB}, because the simultaneous operation of the dual atom interferometer and certain parameters choices allow to engineer suppression ratios for inertial phase shifts and inhomogeneities in the beam splitting wave fronts. A detailed derivation and discussion of error terms for an UFF-test with $^{87}$Rb / $^{85}$Rb in the 10\,m tower in Stanford was reported in~\cite{Hogan08arXiv} and the error budget for a satellite based test can be found in~\cite{Aguilera2014CQG,Schubert13arXiv}. This paper utilizes the same approaches for error assessment and thus focuses on the results. \\
We consider three different scenarios. In the near future, atoms will be dropped from the top chamber, and the scaling factors $k_{Rb}T_{Rb}^{2}=k_{Yb}T_{Yb}^{2}$ will be matched. In this case of matched scaling factors, correlation between the two atom interferometers will then allow to extract the differential phase corresponding to the differential acceleration via ellipse fitting~\cite{Varoquaux2009NJP,Foster2002OL}. The next intermediate step is to use the same free evolution time $T_{Rb}=T_{Yb}$ which mitigates bias terms $\sim kT^{3}$, $\sim kT^{4}$ but requires a more complex read out scheme. Since the scale factors differ now, the correlated signal will not form an ellipse. Restricting phase excursion to below 2$\pi$ still allows the extraction of the differential phase via fitting the Lissajous figure~\cite{Chen2014PRA}. However, the expected vibration noise level is above 2$\pi$. As mentioned earlier this ambiguity may be lifted via correlation with classical sensor mounted in close proximity to the retro reflection mirror as demonstrated for an atom interferometer on a plane~\cite{Geiger2011natcomm} or by adapting the phase extraction algorithms. Finally, the advanced scenario considers launched atoms from the bottom chamber and increased momentum transfers by the beam splitters. A lattice launching technique inside a 10\,m fountain~\cite{Dickerson2013PRL} and high momentum transfer beam splitters~\cite{Chiow2009PRL,Chiow2011PRL} which meet the requirements of this paper were already successfully implemented by other experiments. Requirements for systematics are summed up in table~\ref{tab:error_budget_reqs} and the resulting uncertainties in table~\ref{tab:error_budget_table}. Statistical fluctuations in these parameters are allowed up to the levels reported in table~\ref{tab:statistical_errors_reqs} which implies the errors in table~\ref{tab:statistical_errors_table}. \\
\begin{table}[tb]
 \begin{center}
 \caption{Requirements on noise sources for the dual species atom interferometers in different configurations. All contributions are expected to be uncorrelated. The requirements were set to reach the shot noise limit. Where appropriate values are given as a requirement for a single measurement cycle. (1) Assuming correlation with an additional classical seismometer or advanced data fitting eliminating the $2\pi$ ambiguity.}
 \begin{tabular}{p{0.25\textwidth}p{0.325\textwidth}p{0.325\textwidth}}\hline
   Noise source & Near / intermediate & Advanced \\ \hline
  Shot noise & \multicolumn{2}{c}{\textit{See tab.~\ref{tab:error_budget_reqs} for N, k, and T.}} \\ 
  Beam splitter & \multicolumn{2}{c}{1\,kHz Lorentzian linewidth} \\ 
  Linear vibrations & $10^{-6}\,\mathrm{m\,s}^{-2}\,\mathrm{Hz}^{-1/2}$ & $10^{-6}\,\mathrm{m\,s}^{-2}\,\mathrm{Hz}^{-1/2}$ $^{(1)}$ \\   
  Starting velocity & $\sigma_{v}<0.3$\,mm/s & $\sigma_{v}<3.8\,\mu$m/s \\
  Overlap & $\sigma_{\Delta r}<10\,\mu$m, & $\sigma_{\Delta r}<0.3\,\mu$m, \\
   & $\sigma_{\Delta v}<10\,\mu$m/s & $\sigma_{\Delta v}<0.3\,\mu$m/s \\
  Magnetic fields & $\sigma_{\delta B}<0.5$\,mG/m & $\sigma_{\delta B}<45\,\mu$G/m \\ 
  Wave fronts & \multicolumn{2}{c}{$\sigma_{df}=\sigma_{\Delta z}=100\,\mu$m, jitter telescope \& mirror position 1\,mm} \\
   & \multicolumn{2}{c}{in z-direction (g)} \\
  Mean field & 5\,\% jitter in beam splitting ratio, 20\,\% in atom numbers & 1\,\% jitter in beam splitting ratio, 20\,\% in atom numbers \\
  Cycle times & 11\,s & 12.6\,s \\ \hline 
 \end{tabular}
 \label{tab:statistical_errors_reqs}
 \end{center}
 \end{table}
 
\begin{table}[tb]
 \begin{center}
 \caption{Resulting noise contributions following tab.~\ref{tab:statistical_errors_reqs}. All contributions are expected to be uncorrelated. The requirements were set to reach the shot noise limit. All values are given as the noise of a single measurement.}
 \begin{tabular}{p{0.25\textwidth}p{0.325\textwidth}p{0.325\textwidth}}\hline
   Noise source & Near / intermediate & Advanced \\
   & in $10^{-10}$\,m/s$^{2}$ & in $10^{-11}$\,m/s$^{2}$ \\ \hline
  Shot noise & 4.8 & 1.8 \\ 
  Beam splitter & 2.8 & 1 \\   
  Linear vibrations & 2.8 & 1.8 \\ 
  Overlap & 1 & 0.3 \\ 
  Starting velocity & 0.1 & 0.03 \\
  Magnetic fields & 0.3 & 0.3 \\ 
  Wave fronts & 0.12 / $<0.01$ & $<0.01$ \\
  Mean field & 0.6 & 0.4 \\ \hline 
  Sum & 6.3 & 2.8 \\
  - after 24 h & - 7.1$\cdot$10$^{-2}$ & - 3.4$\cdot$10$^{-2}$ \\ \hline
 \end{tabular}
 \label{tab:statistical_errors_table}
 \end{center}
 \end{table} To engineer a high common mode rejection ratio, the center of mass positions, center of mass velocities, size and expansion ratios of the two atomic species have to be matched. Coupled to gravity gradients and rotations, position and velocity differences in the center of mass positions cause spurious phase shifts in the differential signal. Using trapping frequencies of $2\pi\cdot 500\,$Hz implies a gravitational sag of 1\,$\mu$m which will need to be characterized to $1\,\%$ in the advanced scenario. Due to the lattice launch, we expect a differential velocity of 31\,$\mu$m/s. The corresponding biases will be subtracted from the signal which imposes the requirement of knowing the gravity gradient to 0.1\,\%. This will be measured with the apparatus itself in a gradiometer operation mode. Existing gradiometer experiments reached noise floor of down to $3\cdot10^{-8}\,$s$^{-2}$\,Hz$^{-1/2}$~\cite{McGuirk2002,Rosi2014Nat}. Furthermore, a counter rotation of the retro reflection mirror will reduce the bias due to the earth's rotation~\cite{Dickerson2013PRL}. Additional errors occur if the atoms map different parts of the beam splitter wave fronts to which imperfect collimation or the finite quality of the retro reflection mirror cause inhomogeneities. Commercially available mirrors are rated up to $\lambda/20$ (peak to valley) ~\cite{Fichou} which puts requirements onto the maximum allowable expansion rates. Demonstrated perfomances of lensing $^{87}$Rb atoms to 1\,nK in 3D~\cite{Muntinga2013PRL}, and to 50\,pK in 2D~\cite{Kovachy2014Arx} are sufficient for the experiment.\\
Additional sources for errors are magnetic fields inducing a second order Zeeman shift in the $^{87}$Rb interferometer and the scattering properties of the individual ensembles and the mixture. Suppression of magnetic stray fields with residual rms deviations of $\sim$0.8\,mG inside a three layer 8.8\,m $\mu$-metal shield were demonstrated~\cite{Dickerson2012RSI}. Therefore, additional calibration might be necessary to characterize the magnetic fields to the required level. 


\section{Conclusion and outlook}
\label{outlook}
We presented a novel experimental scheme to test the EEP with two different atomic species, namely ytterbium and rubidium which is in the progress of being set up in Hanover in the new infrastructure of the Hanover Institute of Technology. Using this particular test pair for precision inertial sensing with atom interferometry imposes some challenges which are discussed in this letter together with appropriate specific solutions. Based on the knowledge of this kind of measurement we provide an assessment of the expected performances of the experiment and of the major systematic effects. They should allow to test the E\"otv\"os parameter at a level of \target in the next few years. The work described in this letter is the first step in a complete investigation of inertial sensing with an alkaline earth like element as ytterbium. In the framework of the collaborative research center geo-Q we will investigate possible applications of this technology for geodesy and further ways to improve ground based EEP tests beyond the level of tests with devices employing classical test masses. We expect this work to have a major influence on to the field of fundamental sciences by giving new limits to possible violation scenarios. Moreover, the possibility to investigate interferometric techniques on long time scales with a high repetition rate will benefit atom interferometry experiments in micro-gravity environment or space platforms.

\section*{Acknowledgements}
This work is supported by the DFG in the scope of the SFB geo-Q and will facilitate the major research instrumentation { \it VLBAI-Teststand } applied for at the DFG. The authors would like to also acknowledge the support of the German Space Agency (DLR) with funds provided by the Federal
Ministry of Economic affairs and Energy (BMWi) due to an enactment of the German Bundestag under Grant No. DLR 50WM1131-1137 (project QUANTUS-III). We would like to thank M. Kasevich, J. Hogan and A. Wanner for their help during the planning of the { \it VLBAI-Teststand}. We thank H. Mueller, M. Hohensee, W. Schleich and A. Roura for support concerning the calculation and interpretation of the violation parameters. We thank C. Klempt for fruitful discussions and L. Richardson, P. Berg and E. Wodey for proof reading this document. \\

\section*{References}

\bibliography{VLBAIEPTEST.bbl}

\end{document}